
\magnification=1200
\hsize=15truecm
\vsize=23truecm
\baselineskip 18 truept
\voffset=-0.5 truecm
\parindent=1cm
\overfullrule=0pt

\def\Ai{\hbox{\hbox{${\cal A}$}}\kern-1.9mm{\hbox{${/}$}}}
\def\Vi{\hbox{\hbox{${\cal V}$}}\kern-1.9mm{\hbox{${/}$}}}
\def\Di{\hbox{\hbox{${\cal D}$}}\kern-1.9mm{\hbox{${/}$}}}
\def\lam{\hbox{\hbox{${\lambda}$}}\kern-1.6mm{\hbox{${/}$}}}
\def\D{\hbox{\hbox{${D}$}}\kern-1.9mm{\hbox{${/}$}}}
\def\A{\hbox{\hbox{${A}$}}\kern-1.8mm{\hbox{${/}$}}}
\def\V{\hbox{\hbox{${V}$}}\kern-1.9mm{\hbox{${/}$}}}
\def\parz{\hbox{\hbox{${\partial}$}}\kern-1.7mm{\hbox{${/}$}}}
\def\B{\hbox{\hbox{${B}$}}\kern-1.7mm{\hbox{${/}$}}}
\def\R{\hbox{\hbox{${R}$}}\kern-1.7mm{\hbox{${/}$}}}
\def\si{\hbox{\hbox{${\xi}$}}\kern-1.7mm{\hbox{${/}$}}}

\centerline
{\bf Twistor--like Formulation of the Supermembrane in D=11 $^*$}

\vskip 2truecm

\centerline{\bf P. Pasti and M. Tonin}

\vskip 1truecm

\centerline
{\sl Dipartimento di Fisica ``G. Galilei" -- Universit\`a  di Padova}
\vskip 1truecm
\centerline
{\sl Istituto Nazionale di Fisica Nucleare - Sezione di Padova}
\vskip 0.5truecm
\centerline{\sl Padova (Italy)}

\vskip 2truecm

\noindent
{\bf Abstract.} We propose a new formulation of the D=11
supermembrane theory
that involves commuting spinors (twistor--like variables) and
exhibits a manifest $n$--extended world volume supersymmetry
$(1\leq n\leq 8)$. This supersymmetry replaces $n$ components of the
usual $\kappa$--symmetry. We show that this formulation is classically
equivalent to the standard one.

\vskip 2truecm

\noindent
$^{(*)}$ Supported in part by M.P.I. This work is carried out in the
framework of the European Community Research Programme ``Gauge Theories,
applied supersymmetry and quantum gravity" with a financial contribution
under contract SC1-CT92-D789.

\vskip 1truecm

\noindent
{\bf DFPD/93/TH/07}\hfill {\bf February 1993}

\vfill\eject

\noindent
{\bf 1 - Introduction}

\vskip 0.5truecm

Recently there has been some interest in an alternative approach
$^{[1]-[12]}$ to superparticles and G.S. superstrings in the special
dimensions $D=3,4,6,10$. This approach involves commuting spinors,
i.e. twistor--like variables\footnote{*}{The usefulness of twistors
in string theory has been pointed out by several authors $^{[13]-[16]}$.}
and exhibits both manifest target space supersymmetry and $n$--extended
world line/world sheet supersymmetry $(1\leq n \leq D-2)$. The latter
replaces $n$ components (and therefore provides a
geometrical meaning) of the $\kappa$--symmetry in the standard formulation.
It has been shown by Berkovits $^{[10]}$ that the
$n=2, D=10$ twistor--string
model $^{[4]}$ gives rise to a consistent quantization
of the D=10, G.S. heterotic string with vanishing conformal anomaly.

At least at the classical level, the maximally extended models with
$n=D-2$ are of special interest since in this case the whole
$\kappa$--symmetry is replaced by world line/world sheet supersymmetry.
The correspondence between the critical dimensions $D=3,4,6,10$ and
the division algebras of real, complex, quaternionic and octonionic
numbers respectively has been pointed out several times $^{[15],
[16]}$. However the non associativity of the octonions has represented
an obstacle to get  maximally extended models in
$D=10$. Nevertheless a way to overcome this $D=10$ obstruction has been
found recently for superparticles $^{[7]}$ and heterotic strings
$^{[8],[9]}$. This construction makes use of eight twistors suitably
constrained.
Its geometrical meaning is clear from refs. [6],[7]: the eight
twistors pa\-ra\-me\-tri\-ze the sphere $S^8$, considered as the coset
manifold $SO(9,1)/SO(8)\otimes S^\uparrow(1,1)\times{\bf K}$
where ${\bf K}$ represents the eight conformal boosts.

Unfortunately the $D=10$ twistor--like heterotic string model
is incomplete. Indeed a consistent world sheet supersymmetric
treatment of the heterotic fermions is still lacking for
$n>2$.

Clearly it is worthwhile to extend the twistor approach to
other models where $\kappa$--symmetry is present, as non
heterotic superstrings and supermembranes $^{[17]}$. A formulation of
type II superstrings with $n=1$ world sheet supersymmetry has been
described in ref. [11]. Supermembranes have been considered in a
different but related approach in ref. [12].

The extension of the approach to supermembranes is interesting for
at least two reasons. Firstly,  for
supermembranes the critical dimension
is $D=11$ so that a supermembrane twistor--like
model in $D=11$ would provide an example of the use of commuting
spinors (twistors) in a dimension $(D=11)$ where the
cyclic $\Gamma$--matrix identity (eq. (1.1) below) does not hold
(it is replaced by eq. (1.2) below). Secondly, from such a model
one can get, by dimensional reduction, the twistor--like
action for
type II A superstrings in $D=10$.

In this paper we give a twistor--like formulation of $D=11$
supermembranes. It involves commuting spinors and shows $n$--extended
world volume supersymmetry with $1\leq n \leq 8$.

In sect. 2 we review the main ingredients involved in the twistor--like
approach for the heterotic strings. In sect. 3 we discuss the
constraints and describe the classical twistor--like action for
supermembranes in $D=11$. In sect. 4 we show that this action give
rise to the same field equations of the standard one.

In a forthcoming paper we shall derive from this new supermembrane
formulation the twistor--like action for the $D=10$, type II A superstrings
by performing, as in ref. [18], a simultaneous reduction of one world
volume and one target space dimension.

As for our notations, vector and spinor indices are denoted by
Latin and Greek letters respectively and Capital letters stand for
both kind of indices. Moreover we shall follow the convention of
ref. [9] to write indices of the target space $\underline{{\cal M}}(D|N)$
(of the world manifold ${\cal M}(d|n))$ as underlined (non
underlined) letters. Letters from the beginning of the alphabet
are kept for the tangent spaces.\hfill\break $d$
and $D$  ($n$ and $N$) are
the bosonic (fermionic) dimensions of ${\cal M}(d|n)$ and
${\underline{\cal M}}(D|N)$ respectively. We shall
consider only those dimensions
$D$ where Majorana spinors exist. In these dimensions
$\Gamma^{\underline{a}}_{\underline{\alpha\beta}}=
(C\gamma^{\underline{a}})_{\underline{\alpha\beta}}$ and
$\Gamma^{\underline{a\alpha\beta}}=(\gamma^{\underline{a}}
C^{-1})^{\underline{\alpha\beta}}$ are symmetric
($\gamma^{\underline{a}}$ are Dirac matrices and $C$ is the
charge conjugation matrix in $D$ dimensions). The corresponding
matrices in $d=3$ are denoted by $\sigma^a_{\alpha\beta}$ and
$\sigma^{a\alpha\beta}$. In particular $\sigma_0 =
{1\ 0\choose 0\ 1}$; $\sigma_1 = {0\ 1\choose 1\ 0}$,
$\sigma_2={1\ 0\choose 0\ -1}$ and $\sigma_+={1\over 2} (\sigma_0+
\sigma_2)={1\ 0\choose 0\ 0}$; $\sigma_-={1\over 2}(\sigma_0-
\sigma_2)={0\ 0\choose 0\ 1}$. Indices between round brackets (square
brackets) are symmetrized (antisymmetrized). However the antisymmetric
product of $p\ \Gamma's$ (i.e. $p\ \gamma^{\underline{a}}$ times $C$ or
$C^{-1}$) is denoted by $\Gamma^{\underline{a}_1\cdots\underline{a}_p}$.
The same for $\sigma^{ab}$. In $D=3,4,6,10$ one has the fundamental
cyclic identity

$$
\Gamma_{\underline{a}(\underline{\alpha}\underline{\beta}}
\Gamma^{\underline{a}}_{\underline{\gamma})\underline{\delta}}=0\eqno(1.1)
$$

\noindent
In $D=11$ the cyclic identity is

$$
\Gamma_{\underline{ab}(\underline{\alpha\beta}}
\Gamma^{\underline{b}}_{
\underline{\gamma\delta})}=0\eqno(1.2)
$$

\vskip 0.5truecm

\noindent
{\bf 2 - Twistor--like heterotic strings.}

\vskip 0.5truecm

Before discussing supermembranes, it is convenient to review
the two main ingredients on which is based the twistor--like
formulation of the
G.S. heterotic strings.
These models describe
the embedding of the superworld ${\cal M}(2|n)$ into the target
superspace $\underline{\cal M}(D|2(D-2)$, where $D=3,4,6,10$ and
$1\leq n \leq D-2$.

${\cal M}$ is parametrized
locally by $\zeta^I\equiv(\xi^{(\pm)},\eta^{(q)})$ where $\xi^{(\pm)}$
are the coordinates of the world sheet and $\eta^{(q)}$ are real
Grassman parameters $(q=1,...n)$. The superzweinbeins $e^A
\equiv(e^\pm, e^q)$ define a preferred frame in the cotangent space
of ${\cal M}$, with structure group $SO(1,1)\otimes SO(n)$.
$SO(1,1)$ is the Lorentz group in $d=2$ and $SO(n)$ acts on $e^q$.
The indices $\pm$ and $q$ are rised and lowered with the light--cone
metric and the euclidean metric respectively. $d=e^+ D_+ + e^-
D_- + e^q D_q$ is the differential and $\Delta=e^+\Delta_+ + e^-
\Delta_- + e^q \Delta_q$ is the covariant differential. The torsion
${\cal T}^{\cal A}=\Delta e^{\cal A}$ has the following structure

$$
{\cal T}^+ = 0;\quad
{\cal T}^- = e^q\wedge e_q;\quad
{\cal T}^q = e^+ \wedge e^- {\cal T}^q_{-+}\eqno(2.1)
$$

\noindent
which is compatible with the relevant Bianchi identities.

The target superspace $\underline{\cal M}$ is parametrized locally
by the string supercoordinates $Z^{\underline M} (\zeta)$ which are
world sheet superfields. The tangent space geometry of
$\underline{\cal M}$ is described by the supervielbeins
$E^{\underline A}(Z)$, the Lorentz valued superconnection
$\Omega_{\underline A}^{{\phantom{\underline A}{\underline B}}}(Z)$,
the Lie--G valued
superconnection $A(Z)$ and the two--superform $B(Z)$ with
their curvatures $T^{\underline A}$,
$R_{\underline A}^{{\phantom{\underline A}{\underline B}}}, F$
and $H=dB$. The intrinsic components of these curvatures are restricted
by the SUGRA--SYM constraints

$$
T^{\underline{a}}_{\underline{\alpha\beta}} =
2\Gamma^{\underline{a}}_{\underline{\alpha\beta}};\quad
F_{\underline{\alpha\beta}}=0;\quad
H_{\underline{\alpha\beta\gamma}}=0\eqno(2.2a)
$$

$$
T^{\underline{\alpha}}_{\underline{\beta\gamma}}=0=
T^{\underline{\alpha}}_{\underline{b\beta}}\eqno(2.2b)
$$

$$
H_{\underline{a\beta\gamma}}-\phi(Z)
\Gamma_{\underline{a\beta\gamma}}=0\ ;\
H_{\underline{ab\alpha}} = - {1\over 2}
(\Gamma_{\underline{ab}})_{\underline{\alpha}}\ ^{\underline{\beta}}
\Delta_{\underline{\beta}}\phi\eqno(2.2c)
$$

\noindent
Eq. (2.2b) are conventional constraints and eqs. (2.2c)
follow from
eqs. (2.2a), (2.2b) using the Bianchi identities. $\phi(Z)$ is the dilaton
background superfield.

Moreover $E_A\ ^{\underline{A}}\equiv(E^{\underline{A}}_\pm,
E^{\underline{A}}_q)$ are the intrinsic components of the
pull--back of $E^{\underline{A}}$ on ${\cal M}$ and
$E_A^{{\phantom{A}}{\underline{A}}}
\vert_{\eta^{(p)}=0}\equiv ({\cal E}_\pm^{
\underline{A}}, \lambda^{\underline{A}}_q)$.
Notice that $\lambda^{\underline{\alpha}}_q$ are commuting spinors
(twistors). Twistors enter in these formulations trough the relation

$$
(\lambda_p\Gamma^{\underline{a}} \lambda_q) =
\delta_{pq} {\cal E}_-^{\underline{\alpha}}\eqno(2.3)
$$

\noindent
which, due to the cyclic identity (1.1), implies the Virasoro
constraint ${\cal E}_-^{\underline{a}}{\cal E}_{-\underline{a}}=0$.

The first key ingredient is to implement eq. (2.3) as a world
sheet superfield constraint in order to preserve the world sheet
supersymmetry. This can be done by requiring that the
components of the pull--back
of the vector supervielbeins $E^{\underline{a}}$ along the
world sheet tangent space spinor directions vanish, i.e. by imposing the
following twistor constraint:

$$
E^{\underline{a}}_q = 0 \eqno(2.4)
$$

\noindent
In fact the condition $\Delta_{(p} E^{\underline{a}}_{q)}
\vert_{\eta=0}=0$, that follows from eq. (2.4), reproduces eq. (2.3).
The twistor constraint (2.4) is implemented by the action term

$$
I^{(c)} = \int_{\cal M} P^q_{\underline{a}} E^{\underline{a}}_q
\eqno(2.5)
$$

\noindent
where the lagrangian multipliers $P^q_{\underline{a}}$ are
anticommuting world sheet superfields. For superparticles (not
coupled to a super Maxwell background) this is the whole story
and (the analog of) $I^{(c)}$ represents the full superparticle
action. However for the heterotic strings $I^{(c)}$ must be supplemented
with two further terms, $I^{(B)}$ that involves the two superform
$B={1\over 2}E^{\underline{A}}\wedge E^{\underline{B}}B_{\underline{BA}}$ and
$I^{(h)}$ that describes the heterotic fermions. The problem is to
write $I^{(B)}$ and $I^{(h)}$ without breaking the world sheet
supersymmetry.

In models with $n=1$ this can be done easily $^{[3]}$:

$$
I^{(B)} + I^{(h)} = \int d^2\xi\ d\eta(sdet\ e)
\Bigl[E^{\underline{A}}_+ E^{\underline{B}}_1 B_{\underline{BA}}+i
\psi{\cal D}_1\psi]\Bigr]
$$

\noindent
where $\psi$ is a set of $N_h$ world--sheet  Weyl--Majorana spinors
(heterotic fermions) and ${\cal D}\psi=(\Delta-A)\psi$
(in $D=10$, $N_h=32$).

However for $n>1\ I^{(B)}$ and $I^{(h)}$ cannot be written
so simply as
superspace integrals.

The second key ingredient that allows to write the world sheet
supersymmetric action $I^{(B)}$
is an interesting property $^{[9]}$ carried by $B(Z)$ if
the SUGRA--SYM and twistor constraint holds. Indeed let us consider
the two superform:

$$
\tilde B = B  - {1\over 2n} e^+\wedge e^-
H_{+q}^{\phantom{+q}q}\eqno(2.6)
$$

\noindent
where

$$
H_{+qp}= E^{\underline{A}}_+ E_q^{\underline{B}} E_p^{\underline{C}}
H_{\underline{CBA}}= E_+^{\underline{a}} E_q^{\underline{\beta}}
E_p^{\underline{\gamma}} \phi(Z) \Gamma_{\underline{a\beta\gamma}}
$$

Then, taking into account eqs. (2.1)-(2.4) together with the cyclic
identity (1.1), it is easily to verify that the pull--back of
$d\tilde B$ on the super world sheet vanishes:

$$
d\tilde B|_{\cal M} = 0 \eqno(2.7)
$$

A similar property has been met $^{[19]}$, some years ago, in the
framework of SYM theories. This property, called there Weyl triviality,
was essential to derive the consistent chiral anomaly (the non trivial
BRS cocycle with ghost number one) in SYM theories using the method
of the descend equation. In a similar way the property expressed
by eqs. (2.6), (2.7) allows to get $I^{(B)}$ (the non trivial BRS
cocycle with ghost number zero) and to verify that it is
invariant under world sheet local supersymmetry.
This can be done in two different
but equivalent ways:

\smallskip
\noindent
i) Let  us call ${\cal M}_0$ the slide of ${\cal M}$ at
$\eta^{(q)}= 0 = d\eta^{(q)}$. Then $^{[8]}$

$$
I^{(B)} = \int_{{\cal M}_0} \tilde B = \int_{{\cal M}_0}
e^+\wedge e^- \Bigl[ -{1\over 2n} {\cal E}^{\underline{a}}_+
(\lambda^q\Gamma_{\underline{a}}\lambda_q)\cdot\phi
+ {\cal E}_-^{\underline A} {\cal E}_+^{\underline B}
B_{\underline{BA}}\Bigr]
\eqno(2.8)
$$

\noindent
Under the infinitesimal world sheet supereparametrization

$$
\zeta^I\partial_I\rightarrow\zeta^I\partial_I + \epsilon^A (\xi)
D_A
$$

\noindent
one has

$$
\delta_\epsilon\tilde B|_{\cal M} = (i_\epsilon d\tilde B + d i_\epsilon
\tilde B)|_{\cal M} = d i_\epsilon \tilde B|_{\cal M}
$$

\noindent
where eq. (2.7) has been taken into account. Then

$$
\delta_\epsilon I^{(B)} = 0\eqno(2.9)
$$

Therefore $I^{(B)}$ is invariant
under local supersymmetry even if it is not written
as a full superspace integral. One should notice that, due to
eq. (2.3), eq. (2.8) coincides with the G.S. action (without
heterotic fermions).

\smallskip
\noindent
ii) It follows from eq. (2.7) that locally

$$
\tilde B|_{{\cal M}} = dQ|_{{\cal M}}\eqno(2.10)
$$

\noindent
By imposing eq. (2.10) as a full superspace constraint $^{[9][20]}$,
one has

$$
I^{'(B)} = \int_{\cal M} P^{IJ}
[\tilde B_{JI} - (dQ)_{JI}]\eqno(2.11)
$$

As shown in [9], $I^{'(B)}$ is invariant under a
set of abelian local transformations involving
the superfields $P^{IJ}$. The gauge fixing of these transformations
allows to reduces $I^{'(B)}$ to $I^{(B)}$.

Further abelian transformations involve the lagragian multipliers
$P^q_{\underline{a}}$ in eq. (2.5) and as a consequence of them
the action $I^{(c)} + I^{(B)}$ gives rise to the usual field
equations of the heterotic strings (without heterotic fermions). As
for the heterotic action $I^{(h)}$, a consistent formulation exists
$^{[4]}$ for $n=2$. Unfortunately, up to now, a consistent
supersymmetric version  of $I^{(h)}$ for $n>2$ is still lacking.

\vskip 0.5truecm

\noindent
{\bf 3 - Twistor--like supermembrane: the action}

\vskip 0.5truecm

Twistor--like supermembrane models describe the embedding of the
superworld volume ${\cal M}(3|2n) (1\leq n \leq 8)$
into the target superspace
${\underline{\cal M}}(11|32)$. ${\cal M}(3|2n)$ is
parametrized locally by the supercoordinates $\zeta^M\equiv
(\xi^m, \eta^{q\mu})$ where $m=0,1,2$; $q=1,...n$; $\mu=1,2$ and
$\eta^{q\mu}$ are real Grassman parameters. $e^A=d\zeta^M
e^A_M\equiv (e^a, e^{q\alpha})$ define a local frame in the
cotangent space of ${\cal M}$ and ${\cal T}^A=\Delta e^A$ is the
torsion. $\Delta$ is the covariant differential with respect to the
structure group $SO(2,1)\otimes SO(n)$ and the rheonomic
parametrization of ${\cal T}^A$ is

$$
{\cal T}^a = e^{q\alpha}\wedge e^\beta_q \sigma^a_{\alpha\beta};\quad
{\cal T}^{q\alpha} = e^a\wedge e^{p\beta} {\cal T}^{q\alpha}_{ap\beta}
+ e^a\wedge e^b {\cal T}^{q\alpha}_{ab}\ .\eqno(3.1)
$$

\noindent
${\cal M}(11|32)$ is parametrized locally by the supercoordinates
$Z^{\underline{M}}\equiv (X^{\underline{m}},\theta^{\underline{\mu}})$
$(\underline{m}=0,...10; \underline{\mu}=1,...32)$ which are world
volume superfields. We are interested on supermembranes in $D=11$
supergravity background. $D=11$ supergravity is described by the
supervielbeins $E^{\underline{A}}=dZ^{\underline{M}}
E^{\underline{A}}_{\underline{M}}
(Z)$, the Lorentz superconnection
$\Omega_{\underline{A}}^{\phantom{\underline{A}}{\underline{B}}}
=E^{\underline{C}}
\Omega_{\underline{CA}}\ ^{\underline{B}} (Z) $ and
the 3--superform $B={1\over 3!} E^{\underline{A}}\wedge E^{\underline{B}}\wedge
E^{\underline{C}}B_{\underline{CBA}}(Z)$ with their curvatures i.e.
the torsion $T^{\underline{A}}$, the Lorentz curvature
$R_{\underline{A}}\ ^B$
and the B--curvature $H=dB$. The
intrinsic components of these curvatures are restricted by the SUGRA
constraints:

$$
T^{\underline{a}}_{\underline{\beta\gamma}}=2
\Gamma_{\beta\gamma}^{\underline{a}};\quad
H_{\underline{\alpha\beta\gamma\delta}}=0=
H_{\underline{a\beta\gamma\delta}}\eqno(3.2)
$$

$$
H_{\underline{a b\beta\gamma}}=-{1\over 3}
(\Gamma_{\underline{ab}})_{\underline{\beta\gamma}}\eqno(3.2b)
$$

$$
T^{\underline{\alpha}}_{\underline{\beta\gamma}}=0=
T^{\underline{a}}_{\underline{b\gamma}};\quad
H_{\underline{abc\gamma}}=0\eqno(3.2c)
$$

\noindent
that imply the field equations for the $D=11$, SUGRA.

For the pull--back of $E^{\underline{A}}$ we write

$$
E^{\underline{A}}\Bigl\vert_{\cal M} = e^A E_A^{\underline{A}}
$$

\noindent
and

$$
\left(E_a^{\underline{A}},
E_{q\alpha}^{\underline{A}}\right)_{\eta^{p\mu}=0}=
({\cal E}_a^{\underline{A}},\lambda_{q\alpha}^{\underline{A}}).
$$

\noindent
$\eta_{ab}$ and $\eta_{\underline{ab}}$ are the flat Minkowski
metrics in three and eleven dimensions respectively.

Finally let us recall that the standard supermembrane action is
$^{[17]}$

$$
I_{SM}=\int_{{\cal M}_0} d^3\xi (-det G)^{1/2} +
\int_{{\cal M}_0} E^{\underline{A}}\wedge E^{\underline{B}}\wedge
E^{\underline{C}} B_{\underline{CBA}}\eqno(3.3)
$$

\noindent
where ${\cal M}_0$ is the slide of ${\cal M}$ at $\eta^{
q\mu}=0=d\eta^{q\mu}$ and $G_{mn}$ is the world volume metric induced
by $E_m^{\underline{a}}$:

$$
G_{mn} = E_m^{\underline{a}}\eta_{\underline{ab}}
E_n^{\underline{b}}\eqno(3.4)
$$

\noindent
Like in twistor models for superparticles and heterotic strings,
we now impose the twistor constraint

$$
E_{q\alpha}^{\underline{a}}=0\eqno(3.5)
$$

\noindent
By taking the derivative $\Delta_{p\beta}$ of eq. (3.5) one gets

$$
(E_{q\alpha}\Gamma^{\underline{a}} E_{p\beta})=
\delta_{qp} \sigma^a_{\alpha\beta} E_a^{\underline{a}}\eqno(3.6)
$$

\noindent
Eqs. (3.5), (3.6) give for $\eta^{q\mu}=0$

$$
\lambda_{q\alpha}^{\underline{a }}=0\eqno(3.7a)
$$

$$
(\lambda_{q\alpha}\Gamma^{\underline{a}}
\lambda_{p\beta})=\delta_{qp} \sigma^a_{\alpha\beta}
{\cal E}_a^{\underline{a}}\eqno(3.7b)
$$

\noindent
Before discussing the twistor--like supermembrane action, let us
describe some useful identities that follow essentially from the
cyclic identity (1.2) together with the eq. (3.6). Our first identity
is

$$
{1\over 2}\sigma^{\alpha\beta}_c
(E_{q\alpha}\Gamma_{\underline{ab}} E_{p\beta})
E^{\underline{a}}_a E^{\underline{b}}_b=
\delta_{qp} \epsilon_{abc}\ M\eqno(3.8)
$$

\noindent
where

$$
M=-{1\over 6} \epsilon^{abc} F_c^{\underline{ab}}
E_{a\underline{a}} E_{b\underline{b}}\eqno(3.9)
$$

$$
F^{\underline{ab}}_{c(qp)}={1\over 2}
\sigma_c^{\alpha\beta}
(E_{q\alpha}\Gamma^{\underline{ab}}E_{p\beta})\eqno(3.10)
$$

\noindent
and

$$
F^{\underline{ab}}_c = {1\over n} F^{\underline{ab}}_{c(qp)}
\delta^{qp}
$$

\noindent
First of all let us show that the left hand side of eq. (3.8) is
diagonal in $q$ and $p$. Indeed from eq. (3.6) the l.h.s. of
eq. (3.8) can be written as

$$
M^{qp}_{cab}\equiv {1\over 8} \sigma^{\alpha\beta}_c
\sigma_a^{\gamma\gamma'}\sigma_b^{\delta\delta'}
(E_{q\alpha}\Gamma^{\underline{ab}} E_{p\beta})
(E_{r\gamma}\Gamma^{\underline{a}} E_{r\gamma'})
(E_{r\delta}\Gamma^{\underline{b}} E_{r\delta'})\eqno(3.11)
$$

\noindent
where at least one of the triples $(\beta,\gamma,\gamma')$,
$(\beta,\delta,\delta')$, $(\alpha,\gamma,\gamma')$, $(\alpha,
\delta,\delta')$ is completely symmetric. Suppose that the
triple $(\beta,\gamma,\gamma')$ is symmetric and take $r=p$. Then
from the cyclic identity

$$
(E_{q\alpha} \Gamma_{\underline{ab}}
E_{(p\beta})\ (E_{p\gamma} \Gamma^{\underline{a}}E_{p\gamma')}) =-
(E_{(p\gamma}\Gamma^{\underline{ab}}
E_{p\gamma'})\ (E_{p\beta)}\Gamma_{\underline{a}} E_{q\alpha})\eqno(3.12)
$$

\noindent
and due to eq. (3.6) the r.h.s. of eq. (3.12) vanishes for
$p\not = q$ so that $M^{qp}_{abc}=\delta^{qp} M^{(q)}_{abc}$.
Now let us take eq. (3.11) for $q=p=r$. With the notation that
$V_{\{ab\}}$ represents the components of the tensor $V_{ab}$,
symmetric and traceless with respect to $a$ and $b$, let us
consider
$
M^{(q)}_{\{ca\}b}$. Since $\sigma^{\alpha\beta}_{\{c}
\sigma^{\gamma\gamma'}_{a\}}$ is completely symmetric in
$\alpha,\beta,\gamma$ and $\gamma'$, it follows from the cyclic
identity that $M^{(q)}_{\{ca\}b}$ vanishes. This means that the
$SO(2,1)$\ irreps ``5" and ``3" of $M^{(q)}_{cab}$ vanish so that
$M^{(q)}_{cab}$ is proportional to $\epsilon_{abc}$. Finally  to
prove that $M^{(q)}_{cab}$ is independent from $q$ it is sufficient
to apply the cyclic identity to eq. (3.11), taken for $q=p$ and
$r\not =q$. This complete the proof of eq. (3.8).

A similar argument allows to derive from the cyclic identity the
following interesting relation

$$
\sigma_c^{\alpha\beta}(\Gamma_{\underline{ab}})_{\underline{\alpha\beta}}
E^{\underline{\beta}}_{q\beta}  E^{\underline{a}}_a E^{\underline{b}}_b
\epsilon^{abc} = \sigma^{\alpha\beta}_a
(\Gamma_{\underline{a}})_{\underline{\alpha\beta}}
E^{\underline{\beta}}_{q\beta} E_{\underline{b}b} F^{\underline{ab}}_c
\epsilon^{abc}\eqno(3.13)
$$

\noindent
If we define

$$
R^{\underline{a}a}_{qp} = {1\over 2}
\epsilon^{abc} F^{\underline{ab}}_{b
(qp)} E_{\underline{b}c} \eqno(3.14)
$$

\noindent
eq. (3.8) can be written as

$$
R_{qp,a}^{\underline{a}} E_{\underline{a}b} = \delta_{qp}
\eta_{ab}\ M\eqno(3.15)
$$

\noindent
so that we can put

$$
R^{\underline{a}a}_{qp} = \delta_{qp} M\ E^{\underline{a}}_b
(g^{-1})^{ba} + \overline{R}^{\underline{a}a}_{qp}\eqno(3.16)
$$

\noindent
with

$$
\overline{R}^{\underline{a}a}_{qp} E_{\underline{a}b}=0
$$

\noindent
and $g_{ab}$ is the tangent space world sheet metric induced by
$E^{\underline{a}}_a$ i.e.

$$
g_{ab} = E_a^{\underline{a}} E_{\underline{a}b}\eqno(3.17)
$$

\noindent
We need the technical assumption that this metric (or $G_{mn}$ in
eq. (3.4)) is non degenerate and indeed that its signature is
Minkowskian.

By taking into account eqs. (3.14), (3.16), eq. (3.13) yields

$$\eqalign{
(\sigma_a)^{\alpha\beta} & ((1-\bar\Gamma)\Gamma_{\underline{b}})_{
\underline{\alpha\beta}} E^{\underline{\beta}}_{q\beta}
E^{\underline{b}}_b(g^{-1})^{ba}(-det\ g)^{1/2} = \cr
& (\sigma_a)^{\alpha\beta}(\Gamma_{\underline{b}})_{\underline{\alpha
\beta}} E^{\underline{\beta}}_{p\beta} \Bigl\{
\delta^p_q [(-det\ g)^{1/2}-M] E^{\underline{b}}_b
(g^{-1})^{ba} -{1\over n} \overline{R}_q^{p\underline{b}a}\Bigr\}\cr}
\eqno(3.18)
$$

\noindent
where

$$
(\bar\Gamma)_{\underline{\alpha}}\ ^
{\underline{\beta}}\ =
{\epsilon^{abc} E^{\underline{a}}_a E^{\underline{b}}_b
E_c^{\underline{c}} (\Gamma_{\underline{abc}})_{\underline{\alpha}}\ ^
{\underline{\beta}}
\over 6(-det\ g)^{1/2}}\eqno(3.19)
$$

\noindent
Notice that $\bar\Gamma^2 = 1$ so that

$$
Q_\pm = {1\over 2} (1\pm\bar\Gamma)\eqno(3.20)
$$

\noindent
are orthogonal projectors. Eq. (3.18) implies

$$
M=(-det\ g)^{1/2}\eqno(3.21)
$$

$$
((1+\bar\Gamma)\Gamma_{\underline{b}})
E_{q\beta} \overline{R}_{qp}^{\underline{b}b}
(\sigma_b)^{\beta\alpha}=0\eqno(3.22)
$$

\noindent
Eq. (3.21) is obtained from eq. (3.18) by applying to it
$E_{q\alpha}^{\underline{\gamma}}(1+\bar\Gamma)_{
\underline{\gamma}}\ ^{\underline\alpha}$ and eq. (3.22) is recovered by
acting on eq. (3.18) with the projector $Q_+$ and using eq. (3.21).

Now we are ready to come back to the supermembrane action. Under
the twistor and SUGRA constraints, the three superform B enjoies the
property of Weyl triviality $^{[21]}$. Indeed one can consider
the modified superform

$$
\tilde B=B+{1\over 12n} e^a\wedge e^b\wedge e^c
\sigma^{\alpha\beta}_a E^{q{\underline{A}}}_\alpha
E_{q\beta}^{\underline{B}} E_b^{\underline{C}}
E_c^{\underline{D}}H_{\underline{DCBA}}\eqno(3.23)
$$

\noindent
and  using eqs. (3.1), (3.2), (3.5) (3.6) one can see that the
pull--back on ${\cal M}$ of $d\tilde B$ vanishes

$$
d\tilde B|_{\cal M} =0\eqno(3.24)
$$

\noindent
Then under the infinitesimal superdiffeomorphism
$\zeta^I\partial_I\rightarrow\zeta^I\partial_I +\epsilon^A
(\xi)D_A$ one has

$$
\delta_\epsilon\tilde B|_{\cal M} = di_\epsilon \tilde B|_{\cal M}
$$

\noindent
so that the action

$$
I^{(B)} = \int_{{\cal M}_0} \tilde B \equiv
\int_{{\cal M}_0}
(E^{\underline{A}}\wedge E^{\underline{B}}\wedge
E^{\underline{C}} B_{\underline{CBA}}+{1\over 6}
e^a\wedge e^b\wedge e^c \epsilon_{cba} M)\eqno(3.26)
$$

\noindent
is invariant under local supersymmetry, even if it is not a full
superspace integral (recall that ${\cal M}_0$ is the slide of
${\cal M}$ at $\eta^{q\mu}=0=d\eta^{q\mu})$. To get eq. (3.26)
the twistor and SUGRA constraints have been taken into account and
eq. (3.8) has been used.

In conclusion we propose the following action for the
twistor--like supermembrane

$$
I=I^{(B)} + I^{(c)}\eqno(3.27)
$$

\noindent
where $I^{(c)}$ is given in eq. (3.26) and

$$
I^{(c)} = \int_{\cal M} P^{\alpha q}_{\underline{a}}
E_{q\alpha}^{\underline{a}}\eqno(3.28)
$$

\noindent
implements the twistor constraint (3.5), the superfields
$P^{q\alpha}_{\underline{a}}$ being lagrangian multipliers.

The action $I$ is invariant under
diffeomorphisms and
n--extended local supersymmetry of the world volume.
In addition $I$ is also invariant under the local transformations

$$
\delta P^{\alpha p}_{\underline{a}}=\Delta_{\beta q}
\Lambda^{\{\alpha p,\beta q\}}\eqno(3.29)
$$

\noindent
where the superfields $\Lambda^{\{\alpha p, \beta p\}}$ are
symmetric
with respect to $(\alpha q)$ and
$(\beta p)$and traceless in $q$ and $p$ (i.e. $(\sigma^c)_{\alpha
\beta} \delta_{pq} \Lambda^{\{\alpha q,\beta p\}}=0$).
These transformations are similar to those discussed
in ref. [7], [9] for superparticles and heterotic strings.

Finally one should notice that, as in the case of the heterotic strings,
$I^{(B)}$ can be written as a full superspace integral. Since
locally $\tilde B=dQ$ one has

$$
I^{(B)} = \int_{\cal M} P^{IJK}
\Bigl( \tilde B_{IJK} - (dQ)_{IJK}\Bigr)\eqno(3.30)
$$

\vskip 0.5truecm

\noindent
{\bf 4 - Twistor--like supermembrane: the field equation}

\vskip 0.5truecm

{}From eqs. (3.8), (3.21), the last term in the r.h.s. of eq.
(3.26) can be written as

$$\eqalign{
\int_{{\cal M}_0} e^a\wedge e^b\wedge e^c
F_c^{\underline{ab}} & {\cal E}_{\underline{bb}}
{\cal E}_{\underline{aa}}=\int_{{\cal M}_0} d^3\xi
(-det\ e \cdot det\ g\ \cdot det\ e)^{1/2}=\cr
& = \int_{{\cal M}_0}d^3\xi (-det\ G)^{1/2}\cr}\eqno(4.1)
$$

\noindent
where $G_{mn}$, defined in eq. (3.4), is the world volume metric
induced by $E^{\underline{a}}_m$. Indeed, under eq. (3.5),

$$
e^a_m E_a^{\underline{a}} = e_m^A E_A^{\underline{a}}=
E_m^{\underline{a}}
$$

\noindent
Of course the last equality in eq. (4.1) holds modulo terms
proportional to $\lambda_{q\alpha}^{\underline{\alpha}}$.

\noindent
Therefore, by performing a suitable shift of $P_{\underline{a}}^{q
\alpha}$,  the action I can be rewritten in the form

$$
I= \int_{{\cal M}_0}d^3\xi (-det\ G)^{1/2} +\int_{{\cal M}_0}
e^{\underline{A}}\wedge e^{\underline{B}}\wedge
e^{\underline{C}} B_{\underline{CBA}}+
\int_{\cal M} P_{\underline{a}}^{q\alpha} E^{\underline{a}}_{q
\alpha}\eqno(4.2)
$$

\noindent
(We use the same symbol to denote the superfields $P^{q\alpha}_{
\underline{a}}$ in eq. (3.28) and the shifted ones in eq. (4.2)).

The first two integrals in the r.h.s. of eq. (4.2) reproduce the standard
supermembrane action $I_{SM}$, eq. (3.3).

A perhaps surprising feature of the twistor--like supermembrane is that
the world volume metric $G_{mn}$ induced by the target supervielbeins
$E^{\underline{a}}$ is different from the metric specified
by the  local frame $e^a$, suitable to reveal the hidden world volume
supersymmetry of the model.

In order to show that the action I gives rise to the same field equations of
$I_{SM}$ we need the following

\vskip 0.5truecm

\noindent
{\sl Lemma}. If $V^\alpha_{\underline{a}}$ is a target
space vector--world volume spinor such that

\vskip 0.5truecm

$$
V^\alpha_{\underline{a}} {\cal E}^{\underline{a}}_b = 0\eqno(4.3a)
$$

$$
Q_- V^\alpha_{\underline{a}}\Gamma^{\underline{a}}
\lambda_{q\alpha}=0\eqno(4.3b)
$$

\noindent
for some $q$, then $V^\alpha_{\underline{a}}$ vanishes.

Of course the lemma holds even if $V^{\alpha(n)}_{\underline{a}}$
carries a free index $(n)$, in particular for a target vector--world
volume vector $V^{\alpha\beta}_{\underline{a}}=\sigma^{\alpha\beta}_a
V^a_{\underline{a}}$ such that

$$
V^a_{\underline{a}}{\cal E}_b^{\underline{a}}=0=
Q_- V^{\beta\alpha}_{\underline{a}}\Gamma^{\underline{a}}
\lambda_{q\alpha}.
$$

\noindent
$Q_\pm$ are the projectors defined in eq. (3.20).

Let us call $\hat\Gamma^{\underline{u}} (\underline{u}=1,...8)$
the eight $\Gamma$--matrices that span the eight dimensional
subspace orthogonal to $E_a^{\underline{a}}$.
They anticommute with $\bar\Gamma$. Therefore, by taking into account
eq. (4.3a) and with the notation

$$
\lambda^{(\pm)}_{q\alpha}=Q_\pm \lambda_{q\alpha}
$$

\noindent
Eq. (4.3b) becomes

$$
V^\alpha_{\underline{u}}\hat\Gamma^{\underline{u}}
\lambda^{(+)}_{q\alpha} = 0
$$

\noindent
or,

$$
L_{{\underline{\sigma}\ \alpha}}^{\phantom{\underline{\sigma}\alpha}
{\underline{u}}}
V^\alpha_{\underline{u}}=0\qquad\qquad ({\underline{\sigma}}=1,...,16)
$$

\noindent
where the $16\times 16$ matrix $L$ is
$L_{{\underline{\sigma}\ \alpha}}^{\phantom{\underline{\sigma}\alpha}
{\underline{u}}}
=(\hat\Gamma^{\underline{u}}\lambda^{
(+)}_\alpha)_{\underline{\sigma}}$. To prove the Lemma it is
sufficient to show that the determinant of $L$ is different
from zero. Consider the equation

$$
L_{{\underline{\sigma}\ \alpha}}^{\phantom{\underline{\sigma}\alpha}
{\underline{u}}}
Y^\alpha_{\underline{u}}=0
$$

\noindent
or more explicitly

$$
(\hat Y_1 \lambda^{(+)}_1)_{\underline{\alpha}}
= -  (\hat Y_2 \lambda_2^{(+)})_{\underline{\alpha}}\eqno(4.4)
$$

\noindent
where $Y_{\underline{a}}^\alpha$ are commuting vectors orthogonal to
$E_a^{\underline{a}}$ and $\hat Y_\alpha=
Y_{\underline{u}}^\alpha+
\hat\Gamma^{\underline{u}}$. Eq. (4.4) yields the following identity

$$
(\lambda_1^{(+)} \hat Y_1 \Gamma^{\underline{a}}\hat Y_1
\lambda_1^{(+)}) = (\lambda^{(+)}_2 \hat Y_2 \Gamma^{\underline{a}}
\hat Y_2\lambda^{(+)}_2).
$$

\noindent
which can be rewritten as

$$
(Y_1^{\underline{u}} Y_{1\underline{u}})
E_+^{\underline{a}}=(Y_2^{\underline{u}}
Y_{2\underline{u}}) E^{\underline{a}}_-
$$

\noindent
But $E_+^{\underline{a}}$ and $E_-^{\underline{a}}$ are linearly
independent so that

$$
(Y_1^{\underline{u}} Y_{1\underline{u}})=0=
(Y_2^{\underline{u}} Y_{2\underline{u}})
$$

\noindent
and therefore $Y^\alpha_{\underline{u}}=0$ (in the tangent
subspace orthogonal to $E_a\ ^{\underline{a}}$ the metric is euclidean).
This proves that $det L\not =0$. Then $V_{\underline{u}\alpha}$
vanishes independently of its statistic.

\smallskip

Let us consider at first the model with $n=1$. In this
case the action (4.2) reduces to

$$\eqalign{
I=\int_{{\cal M}_0} & \{ [(-det\ G)^{1/2} + B] + P^a_{\underline{a}}
({\cal E}_a^{\underline{a}}-{1\over 2} \sigma_a^{\alpha\beta}
(\lambda_\alpha \Gamma^{\underline{a}}\lambda_\beta))+\cr
+ & {1\over 2} P^{(0)\alpha}_{\underline{a}}((\sigma^a)_\alpha^\beta
({\cal E}_a \Gamma^{\underline{a}}\lambda_\beta)-3(Y\Gamma^{\underline{a}}
\lambda_\alpha))\}\cr}\eqno(4.6)
$$

\noindent
where

$$\eqalign{
P^a_{\underline{a}} & =\Delta_\alpha(\sigma_a)^\alpha_\beta
P_{\underline{a}}^\beta\Bigl\vert_{\eta^\mu=0}\ ;
P^{(0)\alpha}_{\underline{a}}=
P^\alpha_{\underline{a}}\Bigl\vert_{\eta^\mu=0}\cr
Y^{\underline{A}} & = {1\over 2} \epsilon^{\beta\gamma}\Delta_\beta
E^A_\gamma\Bigl\vert_{\eta^\mu=0}\cr}
$$

\noindent
and $\lambda_\alpha^{\underline{a}}, Y^{\underline{a}}$ have been
eliminated trough their field equations. Then the relevant field
equations are

$$
{\delta I\over\delta Y^{\underline{\alpha}}}\equiv
P^{(0)\alpha}_{\underline{a}}(\Gamma^{\underline{a}}
\lambda_\alpha)_{\underline{\alpha}}
=0\eqno(4.7)
$$

$$
e^I_\gamma {\delta I\over\delta e^I_a} \equiv
P^{(0)\alpha}_{\underline{a}}(\sigma_a\sigma_b)_{\alpha\gamma}
{\cal E}^{\underline{a}b}=0\eqno(4.8)
$$

$$
e^I_b {\delta I\over\delta e^I_a}\equiv P_{\underline{a}a}
E_b^{\underline{a}}+ {1\over 2}
P^{(0)\alpha}_{\underline{a}}(\sigma_a)^\beta_\alpha
(\lambda_\beta \Gamma^{\underline{a}}{\cal E}_b)=0\eqno(4.9)
$$

$$
{\delta I\over\delta\lambda_\alpha^{\underline{\alpha}}}\equiv
P^a_{\underline{a}}(\sigma_a)^{\alpha\beta}
(\Gamma^{\underline{a}}\lambda_\beta)_{\underline{\alpha}}+{1\over 2}
P^{(0)\beta}_{\underline{a}}\Gamma^{\underline{a}}_{\underline{\alpha\beta}}
({\cal E}^{\underline{\beta}}_a(\sigma^a)^\alpha_\beta -3
\delta^\alpha_\beta Y^{\underline{\beta}})=0\eqno(4.10)
$$

$$
E_{\underline{a}}^{\underline{M}}
{\delta I\over\delta Z^{\underline{M}}}\equiv
{\cal L}_{\underline{a}} - \Delta_a P^a_{\underline{a}}=0\eqno(4.11)
$$

$$
E^{\underline{M}}_{\underline{\alpha}}
{\delta I\over\delta Z^{\underline{M}}}\equiv
((1-\bar\Gamma)S)_{\underline{\alpha}}+ {1\over 2} \Delta_a
(P^{(0)\alpha}_{\underline{a}}(\sigma^a)_\alpha^\beta
\Gamma^{\underline{a}}\lambda_\beta)_{\underline{\alpha}}+
P^a_{\underline{a}}(\Gamma^{\underline{a}}{\cal E}_a)_{\underline{\alpha}}
=0\eqno(4.12)
$$

$$
{\delta I\over \delta P^a_{\underline{a}}}\equiv
{\cal E}_a^{\underline{a}}-{1\over 2}
\sigma^{\alpha\beta}_a (\lambda_\alpha \Gamma^{\underline{a}}
\lambda_\beta)=0\eqno(4.13)
$$

$$
{\delta I\over\delta P^{(0)\alpha}_{\underline{a}}}\equiv
(\sigma^a)_\alpha^\beta ({\cal E}_a\Gamma^{\underline{a}}
\lambda_\beta)-(Y \Gamma^{\underline{a}}\lambda_\alpha)=0
\eqno(4.14)
$$

\noindent
where

$$
{\cal L}_{\underline{a}}\equiv
{\delta I_{SM}\over\delta Z^{\underline{M}}}\cdot
E_{\underline{a}}^{\underline{M}}\eqno(4.15)
$$

$$
{\cal L}_{\underline{\alpha}}\equiv
{\delta I_{SM}\over\delta Z^{\underline{M}}}
E^{\underline{M}}_{\underline{\alpha}}=
((1-\bar\Gamma)S)_{\underline{\alpha}}\eqno(4.16)
$$

\noindent
and

$$
S_{\underline{\alpha}}=
({\cal E}^{\underline{a}a}\Gamma_{\underline{a}}
{\cal E}_a)_{\underline{\alpha}}
$$

\noindent
{}From eq. (4.7) one has

$$
P^{(0)\alpha}_{\underline{a}}
(\sigma_b\sigma_a)_{\alpha\beta}
{\cal E}^{\underline{a}b}=0
$$

\noindent
which, together with eq. (4.8), implies

$$
P^{(0)\alpha}_{\underline{a}}{\cal E}^{\underline{a}b}=0\eqno(4.17)
$$

\noindent
Due to eqs. (4.7), (4.17) $P^{(0)\alpha}_{\underline{a}}$ satisfies
the conditions of the Lemma so that $P^{(0)\alpha}_{\underline{a}}=0$.
If $P^{(0)\alpha}_{\underline{a}}$ vanishes, eqs. (4.9), (4.10) show
that $P^a_{\underline{a}}$ too fulfils the conditions of the Lemma
so that $P^a_{\underline{a}}=0$. Then eqs. (4.11), (4.12) become the
standard supermembrane field equations

$$
{\cal L}_{\underline{a}}=0={\cal L}_{\underline{\alpha}}\eqno(4.18)
$$

\noindent
Now let us go back to the general case, $n > 1$.

\noindent
The $Z^{\underline{M}}$ field equations are

$$
{1\over n!}
(\eta^2)^n {\cal L}_{\underline{\alpha}}+
(P_{\underline{a}}^{q\alpha}\Gamma^{\underline{a}}
E_{q\alpha})_{\underline{\alpha}}=0\eqno(4.19)
$$

$$
{1\over n!}
(\eta^2)^n {\cal L}_{\underline{a}}+
\Delta_{q\alpha} P^{q\alpha}_{\underline{a}}=0\eqno(4.20)
$$

\noindent
where

$$
\eta^2={1\over 2} \epsilon_{\alpha\beta}
\eta^{q\alpha}\eta^\beta_q
$$

\noindent
Eq,  (4.20) implies

$$
P_{\underline{a}}^{q\alpha}=\eta^{q\beta}(\eta^2)^{n-1}
P^a_{\underline{a}}(\sigma_\alpha)^\alpha_\beta+
\Delta_{p\beta}\tilde\Lambda_{\underline{a}}^{\{ q\alpha,
p\beta\}}\eqno(4.21)
$$

\noindent
where $\tilde\Lambda_{\underline{a}}^{\{q\alpha,p\beta\}}$ is
symmetric with respect to $(q\alpha)$ and $(p\beta)$
and traceless in $q$ and $p$
However the last term in the r.h.s. of eq. (4.21) can be gauged to
zero using the local invariance, eq. (3.29), so that only the
component $P^a_{\underline{a}}$ of $P^{q\alpha}_{\underline{a}}$
survives. At this point one can repeat the argument given in the case $n=1$
to conclude that also $P^a_{\underline{a}}$ vanishes and
therefore that eqs. (4.18) hold. In conclusion we have shown that the
twistor--like supermembrane action, eq. (3.27) with eqs. (3.26),
(3.28), is classically equivalent to the standard one. Eq. (3.27)
exhibit $n$--extended world volume supersymmetry that replaces $n$
components of the usual $\kappa$--symmetry. Moreover, in our
formulation, conformal invariance is manifest, in agreement with a
recent result $^{[21]}$ where a new conformal invariant formulation
of super $p$--branes has been proposed. This feature is
interesting in view of the problem of the quantization and
renormalization of supermembrane models. It confirms and, in some
sense, explaines the conjecture of ref. [22] where, merely on the basis
of $\kappa$--symmetry, it has been argued that, at the quantum
level, the standard, $D=11$, supermembrane theory should be
renormalizable, despite its lack of conformal invariance.

Another bonus of our formulation is that it provides, by
dimensional reduction, a twistor--like formulation for tipe II A
superstrings, as we shall show elsewhere.

\vskip 0.5truecm

\noindent{\bf References}

\item{[1]}
D.P. Sorokin, V.I. Tkach and D.V. Volkov,
Mod. Phys. Lett. \underbar{A4}, (1989), 901;
\item{} D.P. Sorokin, V.I. Tkach, D.V. Volkov and A.A. Zheltukhin
Phys. Lett. \underbar{216B}, (1989) 302;
\item{} D.P. Sorokin, Fortschr.
Phys. \underbar{38}, (1990) 923;
\item{} D.V. Volkov and A.A. Zheltukhin,
Math. Phys. \underbar{17}, (1989) 141 and  Nucl. Phys.
\underbar{335}, (1990) 723;
\item{}  V.A. Soroka, D.P. Sorokin, V.I.
Tkach and D.V. Volkov,  Int. J. Mod. Phys. \underbar{7A}, (1992) 5977
\item{} A. Pashnev and D.P. Sorokin,
preprint JINR E2-92-27, Dubna.

\item{[2]}
P.S. Howe and P.K. Townsend, Phys. Lett. \underbar{259B}, (1991) 285.
\item{} P.K. Townsend, Phys. Lett. \underbar{261B}, (1991) 65;
\item{} J.P. Gauntlett, Phys. Lett. \underbar{272B}, (1991) 25;
\item{} Y. Eisenberg,  Phys. Lett.
\underbar{276B}, (1992) 325;
\item{} M.S. Plyushchay, Phys. Lett.
\underbar{240B}, (1991) 133;
\item{} F. Delduc and E. Sokatchev,  Class Quantum Grav.
\underbar{9}, (1992) 361;
Phys. Lett. \underbar{262B} (1991) 444:
\item{} N. Berkovits, Nucl. Phys. \underbar{B350}, (1991) 193;
Nucl. Phys. \underbar{B358}, (1991) 169;
\item{} F. Delduc, A.S. Galperin and E. Sokatchev,
Nucl. Phys. \underbar{B368}, (1992) 143;
\item{} F. Delduc, E. Ivanov and E. Sokatchev, Nucl. Phys.
\underbar{B384}, (1992) 334;
\item{} S. Aoyama, P. Pasti and M. Tonin, Phys. Lett.
\underbar{283B}, (1992) 213;
\item{} M. Tonin, Padova preprint DFPD 92/TH/59.

\item{[3]}
N. Berkovits, Phys. Lett. \underbar{232B}, (1989) 184.

\item{[4]}
M. Tonin, Phys. Lett. \underbar{266B}, (1991) 312.

\item{[5]}
E.A. Ivanov and A.A. Kapustnikov, Phys. Lett.
\underbar{267B}, (1991) 175.

\item{[6]}
A.S. Galperin, P.S. Howe and K.S. Stelle, Nucl. Phys.
\underbar{B368}, (1992) 248.

\item{[7]}
A.S. Galperin and E. Sokatchev, Phys. Rev. \underbar{D46},
(1992) 714.

\item{[8]}
M. Tonin, Int. J. Mod. Phys. \underbar{7A}, (1992) 6013.

\item{[9]}
F. Delduc, A.S. Galperin, P. Howe and E. Sokatchev,
Phys. Rev. \underbar{D47} (1992), 578.

\item{[10]}
N. Berkovits, Nucl. Phys. \underbar{B379}, (1992) 96;
\item{} N. Berkovits,
Stony Brook preprint ITP-92-42 (1992);
Phys. Lett. \underbar{300B}, (1992) 53.

\item{[11]}
V. Chikalov and A. Pashnev,
preprint JINR E2-92, Dubna.

\item{[12]}
I.A. Bandos and A.A. Zheltukhin, Phys. Lett. \underbar{B288}
(1992) 77; Theor. Mat. Fis. \underbar{88} (1991).

\item{[13]}
A. Ferber, Nucl. Phys. \underbar{B132}, (1978) 55;
T. Shirafuji,  Prog. Theor. Phys.
\underbar{70}, (1983) 18;
\item{} W. Shaw and L. Hughston in Twistor and Physics eds.
T. Bayley and R. Baston (Cambridge University press) p. 218;
\item{} Y. Eisemberg, S. Salomon, Nucl. Phys. \underbar{B309},
(1988) 709.

\item{[14]}
E. Witten, Nucl. Phys. \underbar{B266}, (1986) 245.

\item{[15]}
A. Bengtsson, I. Bengtsson, M. Cederwall and N. Linden, Phys. Rev.
\underbar{D36}, (1987) 1766;
\item{} I. Bengtsson, M. Cederwall
Nucl. Phys. \underbar{B302}, (1988) 81.

\item{[16]}
D. Farlie and C. Manogue, Phys. Rev. \underbar{D36}, (1987) 575.

\item{[17]}
E. Bergshoeff, E. Sezgin and P.K. Townsend, Phys. Lett.
\underbar{189B}, (1987) 75;
\item{} E. Bergshoeff, E. Sezgin, P.K. Townsend,
Ann. Phys. \underbar{185}, (1988) 30;
\item{} M.J. Duff, Class Quantum Grav. \underbar{5}, (1988) 189.

\item{[18]}
M.J. Duff, P.S. Howe, T. Inami and K.S. Stelle, Phys. Lett.
\underbar{191B}, (1987) 70;
\item{} A. Ach\'ucarro, P. Kapusta and K.S. Stelle,
\underbar{232B}, (1989) 302.

\item{[19]}
L. Bonora, P. Pasti and M. Tonin, Phys. Lett.
\underbar{156B}, (1986) 191;
\item{}  Nucl. Phys. \underbar{B261}, (1985) 241;
\underbar{B286} (1987) 150.

\item{[20]}
J.A. de Azcarraga, J.M. Izquierdo and P.K. Townsend,
Phys. Rev. \underbar{D45}, (1992) 3321;
\item{} P.K. Townsend, Phys.
Lett. \underbar{277B}, (1992) 285.

\item{[21]}
E. Bergshoeff, L.A.J. London and P.K. Townsend, Class Quantum
Grav. \underbar{9}, (1992) 2545.

\item{[22]} F. Paccanoni, P. Pasti and M. Tonin, Mod. Phys. Lett.
\underbar{A4}, (1989) 807.

\bye